\documentstyle[preprint,aps,amstex,amssymb]{revtex}

\begin{document}

\preprint{NUP-A-02-13}

\title{Abelian Projection of Massive $\mbox{\boldmath$SU(2)$}$ Yang-Mills 
Theory}

\vspace{.7in}
\author{Shinichi Deguchi
\footnote{Email address: deguchi@@phys.cst.nihon-u.ac.jp}
and Yousuke Kokubo 
\footnote{Email address: kokubo@@phys.cst.nihon-u.ac.jp}}
\address{Institute of Quantum Science, Nihon University \\
Tokyo 101-8308, Japan} 

\draft

\maketitle


\begin{abstract}
\setlength{\baselineskip}{6mm}
We derive an effective Abelian 
gauge theory (EAGT) of a modified $SU(2)$ Yang-Mills theory.  
The modification is made by explicitly introducing 
mass terms of the off-diagonal gluon fields into pure $SU(2)$ Yang-Mills 
theory, in order that Abelian dominance at a long-distance scale is 
realized in the modified theory. 
In deriving the EAGT, the off-diagonal gluon fields involving 
longitudinal modes are treated as 
fields that produce quantum effects on the diagonal gluon field and 
other fields relevant at a long-distance scale.  
Unlike earlier papers, a necessary gauge fixing is carried out 
without spoiling the global $SU(2)$ gauge symmetry. 
We show that the EAGT allows a composite of the Yukawa and 
the linear potentials which also occurs in an extended dual Abelian 
Higgs model. This composite potential is understood to be 
a static potential between color-electric charges. 
In addition, we point out that the EAGT involves the Skyrme-Faddeev 
model. 

\vspace{4mm}


\end{abstract}



\newpage 

\setlength{\baselineskip}{7mm}

\section{Introduction}

Understanding the color-confinement mechanism based on quantum 
chromodynamics (QCD) is a long-standing subject in particle physics. 
It has been argued that the color-magnetic monopole 
condensation leads to color confinement through the dual Meissner effect 
which is described by the dual Abelian Higgs model or 
the dual Ginzburg-Landau theory \cite{Nambu,Suzuki,SST}. 
To confirm this picture within the framework of QCD, 
it is necessary to realize magnetic monopoles in QCD \cite{'t Hooft} 
and to accept {\em Abelian dominance} 
\cite{Ezawa-Iwazaki,Amemiya-Suganuma} as a fact. 
Here Abelian dominance means that, 
at a long-distance scale, only diagonal gluons dominate, 
while effects of off-diagonal gluons are strongly suppressed. 
When the idea of Abelian dominance was first proposed by 
Ezawa and Iwazaki, it was only a hypothesis 
\cite{Ezawa-Iwazaki}. They conjectured that Abelian 
dominance may be achieved if off-diagonal gluons possess effective 
non-zero mass at a long-distance scale and hence do not propagate at this 
scale. A recent Monte Carlo simulation performed by 
Amemiya and Suganuma shows that, in the maximal Abelian (MA) gauge, 
off-diagonal gluons indeed behave like massive vector fields with 
the effective mass $M_{\rm off} \simeq 1.2 $ GeV \cite{Amemiya-Suganuma}.  
This result strongly supports Ezawa-Iwazaki's conjecture, so that  
the Abelian dominance must be realized at a long-distance scale.

Mass generation of off-diagonal gluons would be a nonperturbative effect 
of QCD at a long-distance scale and should be understood 
within the analytic framework of QCD. In fact, an analytic approach based on 
condensation of the Faddeev-Popov ghosts has been made to explain 
the mass-generation mechanism of off-diagonal gluons 
\cite{Schaden,Kondo-Shinohara 1}. 
This attempt seems to be interesting. However the ghost condensation may 
lead to breaking of the  Becchi-Rouet-Stora-Tyutin (BRST) 
symmetry, which causes the problem of spoiling unitarity.

Although the mass-generation mechanism of off-diagonal gluons is not well 
understood analytically at present, respecting the result of Monte Carlo  
simulation, we accept the mass generation as true in the beginning of 
our discussion without questioning its mechanism. 
Accordingly, in the present letter, 
we explicitly incorporate mass terms of the off-diagonal gluon 
fields into the Yang-Mills (YM) Lagrangian to describe dynamics of 
gluons at a long-distance scale. 
Among the massive YM theories without residual Higgs bosons 
\cite{Deguchi 1}, 
we adopt the Stueckelberg-Kunimasa-Got\={o} (SKG) formalism, 
or the non-Abelian Stueckelberg formalism \cite{Kunimasa-Goto}, 
to deal with massless and massive gluon fields in a gauge-invariant 
manner. In this context, we expect that the mass terms of off-diagonal 
gluon fields are dynamically induced by a nonperturbative effect of 
pure YM theory.

In the SKG formalism, Nambu-Goldstone (NG) scalar fields are introduced 
to identify the longitudinal modes of off-diagonal gluons. 
Since the Lagrangian of the SKG formalism is nonpolynomial in the NG scalar 
fields, there is no hope that the SKG formalism is perturbatively
renormalizable \cite{Shizuya}.   
However this is not a serious problem for our discussion, 
because the SKG formalism is now treated as an effective gauge theory 
at a long-distance scale.

Because of Abelian dominance due to the massive off-diagonal gluons, 
phenomena at a long-distance scale will be described in terms of 
the diagonal gluon fields together with other fields relevant to the 
corresponding scale. 
In accordance with this idea, the present letter attempts to derive 
an effective Abelian gauge theory (EAGT) constructed of these fields 
from the SKG formalism. 
(Some preliminary discussions in this attempt have been reported 
in Ref. \cite{Kokubo-Deguchi}.) 
The procedure of deriving the EAGT is a kind of so-called Abelian projection. 
Recently a fashion of Abelian projection of QCD or YM theory has been 
studied by Kondo \cite{Kondo 1,Kondo 2} and by some other authors 
\cite{Reinhardt,Kondo-Shinohara 2,Freire}. 
In their procedure, the maximally Abelian (MA) gauge condition 
is imposed on QCD before performing the Abelian projection. 
Unfortunately, the {\em global} color gauge symmetry of QCD is spoiled 
at this stage, so that the conservation of color Noether current is no 
longer guaranteed. 
In contrast, in our procedure, we do not put the MA gauge condition 
by hand. Instead, a corresponding condition is obtained as the  
Euler-Lagrange equation for the NG scalar fields. 
In connection with this fact the global color gauge symmetry 
is maintained in the EAGT in a nonlinear way, provided that 
an appropriate gauge fixing term is 
adopted.\footnote{\setlength{\baselineskip}{6mm} 
A gauge-covariant formulation of Abelian projection has been considered 
with introducing auxiliary Higgs-like fields
\cite{Brower,Ichie-Suganuma}. Unlike this approach, 
the present letter treats the NG scalar fields (which are Higgs-like fields 
represented in a nonlinear way) as dynamical fields 
occurring in the Lagrangian.} 
As a result, a conserved color charge is well defined in the EAGT. 
The presence of conserved color charge will be important for combining  
our discussion with other attempts to investigate color confinement.

The Abelian projection of the SKG formalism is performed in the following 
manner: First, beginning with an effective action in the SKG formalism, 
we evaluate the path-integral over the NG scalar fields 
around their classical configuration by using a semi-classical method. 
After that, similar to earlier papers 
\cite{Kondo 1,Reinhardt,Kondo-Shinohara 2,Kondo 2,Freire}, 
we carry out the path-integration over the off-diagonal gluon 
fields with the aid of auxiliary fields introduced in this stage. 
The resulting effective action that defines the EAGT is written in terms 
of the diagonal gluon fields, the auxiliary fields and 
the classical configuration of NG scalar fields. 
This action includes terms analogous to those occurring in a dual form 
of the extended dual Abelian Higgs model (EDAHM) 
in the London limit \cite{BBDV,Antonov-Ebert}. 
With such terms in hand, following a procedure discussed in 
Ref. \cite{Deguchi-Kokubo}, 
we show that the EAGT allows a composite of the Yukawa potential and 
the linearly rising potential. 
We point out that the composite potential 
represents interaction between color-electric charges consisting of 
the classical NG scalar fields. 
We also show that the EAGT involves the Skyrme-Faddeev model.

\section{Stueckelberg-Kunimasa-Got\={o} formalism}

For simplicity, in the present letter, we restrict our discussion to 
the case of $SU(2)$ color gauge symmetry. 
Let ${A_{\mu}}^{B}$ $(B=1, 2, 3)$ be the Yang-Mills (YM) fields, 
or the gluon fields, and $\phi^{i}$ $(i=1, 2)$ be the (dimensionless) 
Nambu-Goldstone (NG) scalar fields that 
form a set of coordinates of the coset space $SU(2)/U(1)$.  
We begin with the Stueckelberg-Kunimasa-Got\={o} (SKG) formalism 
\cite{Kunimasa-Goto} characterized by the Lagrangian 
${\cal L}_{\rm SKG}={\cal L}_{\rm YM}+{\cal L}_{\phi}$ with  
\begin{eqnarray}
{\cal L}_{\rm YM} &=& -{1\over{4g_{0}^{2}}} F_{\mu\nu}{}^{B} F^{\mu\nu B} \,, 
\label{1}
\\ 
{\cal L}_{\phi} &=& {m_{0}^{2}\over{2g_{0}^{2}}} \, g_{ij}(\phi)
{\cal D}_{\mu}\phi^{i} {\cal D}^{\mu}\phi^{j} \,, 
\label{2}
\end{eqnarray}
%
where $F_{\mu\nu}{}^{B}\equiv \partial_{\mu}A_{\nu}{}^{B}
-\partial_{\nu}A_{\mu}{}^{B}-\epsilon^{BCD}A_{\mu}{}^{C} A_{\nu}{}^{D}$ and  
${\cal D}_{\mu}\phi^{i}\equiv \partial_{\mu}\phi^{i}
+A_{\mu}{}^{B}K^{Bi}(\phi)$ \cite{Deguchi 2}.   
Here $g_{0}$ is a (bare) coupling constant, 
$m_{0}$ is a constant with dimension of mass, 
$g_{ij}(\phi)$ is a metric on the coset space $SU(2)/U(1)$, and 
$K^{Bi}(\phi)$ are Killing vectors on $SU(2)/U(1)$. 
(The convention for the signature of space-time metric is $(+, -, -, -)$.)
The Lagrangian ${\cal L}_{\rm SKG}$ remains invariant under the gauge 
transformation 
\begin{gather}
\begin{split}
\delta A_{\mu}{}^{B} &= {\cal D}_{\mu}{}^{BC}\lambda^{C} \equiv  
(\delta^{BC}\partial_{\mu}+\epsilon^{BCD}A_{\mu}{}^{D})\lambda^{C} \,, 
\\
\delta\phi^{i} &= -\lambda^{B}K^{Bi}(\phi) \,, 
\label{3} 
\end{split}
\end{gather}
%
with $SU(2)$ gauge parameters $\lambda^{B}$. 
The metric $g_{ij}(\phi)$ is written as 
$g_{ij}(\phi)=e_{i}{}^{b}(\phi)e_{j}{}^{b}(\phi)$ $(b=1,2)$ in terms of 
the zweibein $e_{i}{}^{b}(\phi)$ that is defined  by 
$e_{i}{}^{b}(\phi)T^{b}+e_{i}{}^{3}(\phi)T^{3}=e_{i}{}^{B}(\phi)T^{B}
\equiv -iv^{-1}(\phi)\partial_{i}v(\phi)$ with  
coset representatives $v(\phi)\, [ \in SU(2)\,]$,  
where $\partial_{i}\equiv \partial/\partial\phi^{i}$ 
and $T^{B}\equiv{1\over2}\sigma^{B}$ 
($\sigma^{B}$ denote the Pauli matrices). 
Under the left action of $g\, [ \in SU(2)\,]$, 
the coset representatives transform as 
$gv(\phi)=v(\phi')h(\phi, g)$  $[\,h \in U(1)\,]$; 
in a sense, $v(\phi)$ ^^ ^^ convert" 
$SU(2)$ gauge transformations into corresponding $U(1)$ gauge 
transformations. 
If the infinitesimal form $g=1-i\lambda^{B}T^{B}$ is chosen as $g$,  
then $\phi'{}^{i}$ and $h$ take the 
following forms: $\phi'{}^{i}=\phi^{i}-\lambda^{B}K^{Bi}(\phi)$, 
$h(\phi, g)=1-i\hat{\lambda} T^{3}$. 
Here $\hat{\lambda}$ is a $U(1)$ gauge parameter defined 
by $\hat{\lambda} \equiv \lambda^{B}\varOmega^{B}(\phi)$  
with $\varOmega^{B}(\phi)$ being functions of $\phi^{i}$ 
called $H$-compensators \cite{Nieu}. 
(Unlike $\lambda^{3}$, $\hat{\lambda}$ is not a mere gauge 
parameter of the $U(1)$ subgroup of the gauge group SU(2).) 
From the transformation rule of $v(\phi)$, it follows that 
$K^{Bi}(\phi)e_{i}{}^{a}(\phi)={\Bbb D}^{Ba}(v(\phi))$. 
Here ${\Bbb D}$ denotes a matrix in the adjoint representation of 
$SU(2)$, defined by $g^{-1}T^{B}g={\Bbb D}^{BC}(g)T^{C}$. 
In addition, the transformation rule of $v(\phi)$ leads to 
$[K^{B}, K^{C}]=\epsilon^{BCD}K^{D}$ with 
$K^{B}\equiv K^{Bi}(\phi)\partial_{i}\,$.  
It is easy to show that the inverse metric 
$g^{ij}(\phi)=e^{bi}(\phi)e^{bj}(\phi)$, 
with the inverse zweibein $e^{bi}(\phi)$, 
can be written as $g^{ij}(\phi)=K^{Bi}(\phi)K^{Bj}(\phi)$  
and so the Lie derivatives of $g^{ij}(\phi)$ along $K^{B}$ vanish. 
The Maurer-Cartan formula for  
$e_{\mu}{}^{B}(\phi)\equiv \partial_{\mu}\phi^{i} e_{i}{}^{B}(\phi)$ 
now reads 
\begin{eqnarray}
\partial_{\mu}e_{\nu}{}^{B}-\partial_{\nu}e_{\mu}{}^{B}
-\epsilon^{BCD}e_{\mu}{}^{C} e_{\nu}{}^{D} =\delta^{B3} \varSigma_{\mu\nu}\,,
\label{4}
\end{eqnarray}
%
where $\varSigma_{\mu\nu}$ is given by 
$\varSigma_{\mu\nu}(\phi)T^{3}\equiv 
-iv^{-1}(\phi)[\partial_{\mu}, \partial_{\nu}] v(\phi)$. 
This term remains non-vanishing owing to a Dirac string singularity  
characterized by the first homotopy group $\pi_{1}(U(1))={\Bbb Z}$ 
\cite{SST,Wiese}. 

The Lagrangian ${\cal L}_{\phi}$ can be expressed as 
${\cal L}_{\phi}={1\over2}
(m_{0}/g_{0})^{2}\widehat{A}_{\mu}{}^{b}\widehat{A}^{\mu b}$ 
in terms of the YM fields in the unitary gauge, 
$\widehat{A}_{\mu}{}^{B}
\equiv A_{\mu}{}^{C} {\Bbb D}^{CB}(v(\phi))+ e_{\mu}{}^{B}(\phi)$. 
Thus, in the unitary gauge, ${\cal L}_{\phi}$ takes the form of a mass 
term for $\widehat{A}_{\mu}{}^{b}$. 
The gauge transformation rules of $\widehat{A}_{\mu}{}^{B}$ $(B=b,3)$ 
are found from Eq. (\ref{3}) to be 
\begin{eqnarray}
\delta\widehat{A}_{\mu}{}^{b} = -\epsilon^{bc3}\widehat{A}_{\mu}{}^{c}
\hat{\lambda} \,, 
\qquad 
\delta\widehat{A}_{\mu}{}^{3} = \partial_{\mu}
\hat{\lambda} \,, 
\label{5}
\end{eqnarray}
%
with which we can easily check the gauge invariance of ${\cal L}_{\phi}$.   
We now see that the off-diagonal gluon fields 
$\widehat{A}_{\mu}{}^{b}$ behave like massive charged matter fields with 
the mass $m_{0}$, while the diagonal gluon field  
$\widehat{A}_{\mu}{}^{3}$ behaves like a massless $U(1)$ gauge field. 
For this reason, it is quite natural to interpret the unitary gauge as 
the maximally Abelian (MA) gauge discussed in earlier papers (see, e.g., 
Refs. \cite{Amemiya-Suganuma,Kondo 1,Reinhardt,Ichie-Suganuma}).  
Remarkably the Euler-Lagrange equation for $\phi^{i}$ can be written 
in the form of the MA gauge condition for $\widehat{A}_{\mu}{}^{b}\,$:  
\begin{eqnarray}
\widehat{D}^{\mu bc} \widehat{A}_{\mu}{}^{c}\equiv 
(\delta^{bc}\partial^{\mu}+\epsilon^{bc3}\widehat{A}^{\mu3})
\widehat{A}_{\mu}{}^{c}=0 \,. 
\label{6}
\end{eqnarray}
%
This fact also ensures that $\widehat{A}_{\mu}{}^{B}$ would be 
the YM fields in the MA gauge. 
In the earlier papers, a similar equation for $A_{\mu}{}^{b}$, 
not for $\widehat{A}_{\mu}{}^{b}$, has been put by hand as  
the MA gauge condition, 
whereas in the present letter the above equation has been derived as 
the Euler-Lagrange equation for $\phi^{i}$.

Carrying out the Wick rotation $x^{0}\rightarrow x^{0}=-ix^{4}$ 
$(x^{4}\in{\Bbb R}\,)$ in Eqs. (\ref{1}) and (\ref{2}), 
let us consider the following effective action in Euclidean space-time: 
\begin{eqnarray}
W = \ln \int {\frak D}{\cal M} \exp \! \left[ \, \int d^{4}x 
({\cal L}_{\rm YM}+{\cal L}_{\phi}+{\cal L}_{\rm GF}) \right] 
\label{7}
\end{eqnarray}
%
with ${\frak D}{\cal M}\equiv
{\frak D}A_{\mu}{}^{B} \,{\frak D}\phi^{i} \,{\frak D}c^{B} \,
{\frak D}\bar{c}^{B} \,{\frak D}b^{B}$. 
Here $c^{B}$ and $\bar{c}^{B}$ are Faddeev-Popov ghost fields, 
$b^{B}$ are Nakanishi-Lautrup fields, and ${\cal L}_{\rm GF}$ stands for 
a gauge fixing term (involving ghost terms) that is introduced to break 
the local $SU(2)$ gauge invariance of ${\cal L}_{\rm SKG}$. 
The Becchi-Rouet-Stora-Tyutin (BRST) transformation 
$\boldsymbol{\delta}$ is defined, as usual, by 
\begin{alignat}{2}
\boldsymbol{\delta} A_{\mu}{}^{B} &={\cal D}_{\mu}{}^{BC}c^{C}\,, &\;
\boldsymbol{\delta} \phi^{i} &=-c^{B}K^{Bi}(\phi) \,, 
\nonumber \\ 
\boldsymbol{\delta} c^{B} &={1\over2}\epsilon^{BCD} c^{C}c^{D} \,, 
\label{8}
\\ 
\boldsymbol{\delta} \bar{c}^{B} &=ib^{B}\,, &\;
\boldsymbol{\delta} b^{B} &=0 
\nonumber 
\end{alignat}
%
to satisfy the nilpotency property 
$\boldsymbol{\delta}^{2}=0$. 
We suppose that ${\cal L}_{\rm GF}$ contains no $\phi^{i}$ but may 
contain the ^^ ^^ classical" fields $\phi_{0}{}^{i}$ introduced 
in the following.

\section{Path-integration over the Nambu-Goldstone scalar fields}

We first evaluate the path-integral over $\phi^{i}$ in Eq. (\ref{7}) 
in the one-loop approximation, which can be done covariantly by 
employing the geometrical method proposed by Honerkamp \cite{Hon}. 
In this method, the action $\int d^{4}x {\cal L}_{\phi}$ in the exponent of  
Eq. (\ref{7}) is expanded about a solution 
$\phi_{0}{}^{i}$ of Eq. (\ref{6}) in powers of new integration variables 
$\sigma^{i}$. Here $\sigma^{i}$ are understood to form the tangent vector 
at $\phi_{0}{}^{i}$ to the geodesic 
on $SU(2)/U(1)$ that runs from $\phi_{0}{}^{i}$ to the original 
integration variables $\phi^{i}$. 
Accordingly, the gauge-invariant measure 
${\frak D}_{g} \sigma^{i}\equiv 
({\rm det}[g_{ij}(\phi_{0})])^{1/2} {\frak D}\sigma^{i}$ 
is chosen as ${\frak D}\phi^{i}$. 
After ignoring the third and higher order terms in $\sigma^{i}$, 
the integration over $\sigma^{i}$ in Eq. (\ref{7}) can be performed to get 
the determinant of a certain Laplace-type operator containing 
$A_{\mu}{}^{B}$ and $\phi_{0}{}^{i}$. 
Applying the heat kernel method to the evaluation of the determinant, 
we obtain a one-loop effective action. Addition of this to 
the classical action at $\phi_{0}{}^{i}$ leads to 
$\int d^{4}x \widetilde{\cal L}_{\phi_{0}} 
\approx \ln \int {\frak D}_{g} \sigma^{i} 
\exp \! \big[ \int d^{4}x {\cal L}_{\phi} \big]+\ln N_{0}$, 
where $N_{0}$ is a normalization constant and $\approx$ stands for 
the one-loop approximation.  
(The $N_{l}$ $(l=1,\dots ,6)$ that appear in the following denote 
normalization constants.) 
By using the Maurer-Cartan formula in Eq. (\ref{4}) 
at $\phi^{i}=\phi_{0}{}^{i}$, 
$\,\widetilde{\cal L}_{\phi_{0}}$ is written in terms of  
$\widetilde{A}_{\mu}{}^{B} \equiv 
\widehat{A}_{\mu}{}^{B}|_{\phi=\phi_{0}}=
A_{\mu}{}^{C} {\Bbb D}^{CB}(v(\phi_{0}))+ e_{\mu}{}^{B}(\phi_{0})$ 
as 
\begin{eqnarray} 
\widetilde{\cal L}_{\phi_{0}}&=&\frac{\varLambda^{4}}{32\pi^{2}}
+\!\bigg( \frac{m_{0}^{2}}{2g_{0}^{2}}-\frac{\varLambda^{2}}{32\pi^{2}} \bigg)
{\cal S}
\nonumber \\ 
& &
-{1\over{16\pi^{2}}} \ln \frac{\mu}{\varLambda}
\bigg( \! -{1\over6}{\cal F}_{\mu\nu}{\cal F}^{\mu\nu} 
+{1\over8}(1+3k_{1}){\cal S}^{2} 
\nonumber \\
& &
-{k_{1}\over2}{\cal A}_{\mu\nu} {\cal A}^{\mu\nu} 
+{1\over2}(1-k_{1}){\cal T}_{\mu\nu}{\cal T}^{\mu\nu} \bigg) 
+O_{1}(\varLambda^{-2}) \,,
\label{9}
\end{eqnarray}
%
where ${\cal F}_{\mu\nu}\equiv 
\partial_{\mu} \widetilde{A}_{\nu}{}^{3}
-\partial_{\nu} \widetilde{A}_{\mu}{}^{3}-\varSigma_{\mu\nu}(\phi_{0})$, 
${\cal S}\equiv \widetilde{A}_{\mu}{}^{b} \widetilde{A}^{\mu b}$, 
${\cal A}_{\mu\nu}\equiv -\epsilon^{3bc} 
\widetilde{A}_{\mu}{}^{b} \widetilde{A}_{\nu}{}^{c}$,    
and ${\cal T}_{\mu\nu}\equiv 
\widetilde{A}_{\mu}{}^{b} \widetilde{A}_{\nu}{}^{b}
-{1\over4}\eta_{\mu\nu} {\cal S}$, with $\eta_{\mu\nu}\equiv -\delta_{\mu\nu}$ 
the metric on Euclidean space-time. 
In deriving Eq. (\ref{9}), a mass scale $\mu$ and an ultraviolet 
cutoff $\varLambda$ have been introduced to make 
$\widetilde{\cal L}_{\phi_{0}}$ a meaningful expression. 
In addition, a dimensionless constant $k_{1}$ has been 
introduced to indicate an arbitrariness of expression due to the identity 
${\cal A}_{\mu\nu} {\cal A}^{\mu\nu}={3\over4}{\cal S}^{2}
-{\cal T}_{\mu\nu} {\cal T}^{\mu\nu}$. 
At $\phi^{i}=\phi_{0}{}^{i}$ the gauge transformation rules in Eq. (\ref{5})  
read 
\setcounter{enumi}{\value{equation}}
\addtocounter{enumi}{1}
\renewcommand{\theequation}{\theenumi\alph{equation}}
\setcounter{equation}{0}
\begin{align}
\delta\widetilde{A}_{\mu}{}^{b} &= 
-\epsilon^{bc3}\widetilde{A}_{\mu}{}^{c} \tilde{\lambda} \,, 
\label{10a}
\\ 
\delta\widetilde{A}_{\mu}{}^{3} &= \partial_{\mu}\tilde{\lambda} \,, 
\label{10b}
\end{align}
%
with $\tilde{\lambda} \equiv \lambda^{B}\varOmega^{B}(\phi_{0})$. 
As can be seen from Eq. (\ref{10a}), ${\cal S}$, ${\cal A}_{\mu\nu}$ and 
${\cal T}_{\mu\nu}$ are gauge invariant. 
In Eq. (\ref{9}), 
${\cal L}_{\phi_{0}}\equiv{1\over2}
(m_{0}/g_{0})^{2}{\cal S}$ is the only classical term.

\section{Auxiliary fields}

In terms of $\widetilde{A}_{\mu}{}^{B}$, 
the Lagrangian ${\cal L}_{\rm YM}$ can be written as  
\renewcommand{\theequation}{\arabic{equation}}
\setcounter{equation}{\value{enumi}}
\begin{eqnarray}
{\cal L}_{\rm YM}&=&-\frac{1}{4g_{0}^{2}}
\bigg[ {\cal F}_{\mu\nu}{\cal F}^{\mu\nu} +2{\cal F}_{\mu\nu}{\cal A}^{\mu\nu} 
+k_{2} {\cal A}_{\mu\nu} {\cal A}^{\mu\nu} 
\nonumber \\ 
& &+(1-k_{2}) \bigg({3\over4}{\cal S}^{2} 
-{\cal T}_{\mu\nu} {\cal T}^{\mu\nu} \bigg) \!
+\widetilde{F}_{\mu\nu}{}^{b} \widetilde{F}^{\mu\nu b} \bigg] \,,
\label{11}
\end{eqnarray}
%
where $k_{2}$ is a constant similar to $k_{1}$, and  
$\widetilde{F}_{\mu\nu}{}^{b}$ is the field strength of 
$\widetilde{A}_{\mu}{}^{b}$, expressed as 
$\widetilde{F}_{\mu\nu}{}^{b}
=\widetilde{D}_{\mu}{}^{bc} \widetilde{A}_{\nu}{}^{c} 
-\widetilde{D}_{\nu}{}^{bc} \widetilde{A}_{\mu}{}^{c}$ with 
$\widetilde{D}_{\mu}{}^{bc} \equiv 
\delta^{bc}\partial_{\mu}+\epsilon^{bc3}\widetilde{A}_{\mu}{}^{3}$.

In order to derive an EAGT involving $\widetilde{A}_{\mu}{}^{3}$,  
we need to evaluate the path-integral over $\widetilde{A}_{\mu}{}^{b}$  
in the effective action
\begin{eqnarray}
W' &=& \ln \int {\frak D}{\cal M'} \exp \! \left[\, \int d^{4}x 
({\cal L}_{\rm YM}+\widetilde{\cal L}_{\phi_{0}}+{\cal L}_{\rm GF}) 
\right] 
\nonumber \\ 
(&\approx& W+\ln N_{0}\,) 
\label{12}
\end{eqnarray}
%
with ${\frak D}{\cal M}' \equiv
{\frak D}{\widetilde A}_{\mu}{}^{B} \,{\frak D}c^{B} \,
{\frak D}\bar{c}^{B} \,{\frak D}b^{B}$. 
Here the gauge invariance of the path-integral measure 
${\frak D}A_{\mu}{}^{B}$, i.e., 
${\frak D}A_{\mu}{}^{B}={\frak D}{\widetilde A}_{\mu}{}^{B}$, has been used. 
In order to perform the integration over 
${\widetilde A}_{\mu}{}^{b}$ exactly, we introduce the following 
auxiliary fields: 
an antisymmetric tensor field $B_{\mu\nu}$, a scalar field $\varPhi$ 
and a traceless symmetric tensor field $h_{\mu\nu}$ that are 
associated with ${\cal A}_{\mu\nu}$, ${\cal S}$ and ${\cal T}_{\mu\nu}$, 
respectively.\footnote{\setlength{\baselineskip}{6mm}
In the case of pure $SU(2)$ YM theory without  
the scalar fields $\phi^{i}$, we can exactly carry out the integration 
over ${\widetilde A}_{\mu}{}^{B}$ with the aid of $B_{\mu\nu}$ 
alone \cite{Kondo 1,Reinhardt,Freire}. 
In the present case involving $\phi^{i}$, the introduction of 
$\varPhi$ and/or $h_{\mu\nu}$ is essential for exactly carrying out 
the integration over ${\widetilde A}_{\mu}{}^{B}$.}
With the aid of these new fields, 
$\exp \! \big[ \int d^{4}x ({\cal L}_{\rm YM}
+\widetilde{\cal L}_{\phi_{0}}) \big]$ 
can be expressed as 
$N_{1} \int {\frak D}B_{\mu\nu}\, {\frak D}\varPhi\, {\frak D}h_{\mu\nu} 
\exp \! \big[ \int d^{4}x {\cal L}_{1} \big]$ with 
\begin{eqnarray}
{\cal L}_{1}&=& \frac{\varLambda^{4}}{32\pi^{2}}
-{1\over4}
\bigg(\frac{1}{g_{0}^{2}}-\frac{1}{24\pi^{2}} \ln \frac{\mu}{\varLambda}
-\frac{\kappa^{2}}{q_{1}} \bigg) 
{\cal F}_{\mu\nu}{\cal F}^{\mu\nu} 
+\bigg( \frac{m_{0}^{2}}{2g_{0}^{2}}-\frac{\varLambda^{2}}{32\pi^{2}}
 \bigg){\cal S} 
\nonumber \\ 
& & -{1\over2}\bigg( \frac{1}{g_{0}^{2}}
-\frac{\kappa}{q_{1}}\bigg)
{\cal F}_{\mu\nu}{\cal A}^{\mu\nu} 
+{i\over2} B_{\mu\nu}(\kappa {\cal F}^{\mu\nu}+{\cal A}^{\mu\nu})
-\frac{q_{1}}{4} B_{\mu\nu}B^{\mu\nu} 
\nonumber \\ 
& &+{i\over2}\, \varPhi{\cal S}
-\frac{q_{2}}{4} \,\varPhi^{2} 
+{i\over2} h_{\mu\nu}{\cal T}^{\mu\nu}
-\frac{q_{3}}{4} h_{\mu\nu}h^{\mu\nu} 
-\frac{1}{4g_{0}^{2}} \widetilde{F}_{\mu\nu}{}^{b} 
\widetilde{F}^{\mu\nu b} 
\nonumber \\ 
& &+O_{1}(\varLambda^{-2}) \,.
\label{13}
\end{eqnarray}
%
Here $\kappa$ is an arbitrary dimensionless constant, 
and $q_{1}$, $q_{2}$ and $q_{3}$ are defined by 
\begin{eqnarray}
q_{1}^{-1}&\equiv&
\frac{k_{2}}{g_{0}^{2}}-\frac{k_{1}}{8\pi^{2}} \ln\frac{\mu}{\varLambda} \,,
\label{14}
\\
q_{2}^{-1}&\equiv&
\frac{3(1-k_{2})}{4g_{0}^{2}}
+\frac{1+3k_{1}}{32\pi^{2}} \ln\frac{\mu}{\varLambda} \,,
\label{15}
\\
q_{3}^{-1}&\equiv&
\frac{k_{2}-1}{g_{0}^{2}}
-\frac{k_{1}-1}{8\pi^{2}} \ln\frac{\mu}{\varLambda} \,.
\label{16}
\end{eqnarray}
%
Taking into account the gauge invariance of ${\cal F}_{\mu\nu}$, 
${\cal A}_{\mu\nu}$, ${\cal S}$ and ${\cal T}_{\mu\nu}$,  
we impose the gauge transformation rules 
$\delta B_{\mu\nu} = \delta\,\varPhi = \delta h_{\mu\nu}=0$ on 
the auxiliary fields. Then ${\cal L}_{1}$ is obviously gauge invariant.  
The BRST invariance of ${\cal L}_{1}$ is guaranteed with 
$\boldsymbol{\delta}B_{\mu\nu}= \boldsymbol{\delta}\,\varPhi 
=\boldsymbol{\delta}h_{\mu\nu}=0$. 
The effective action we are now concerned with is thus 
\begin{eqnarray}
W'' &=& \ln \int {\frak D}{\cal M}'' \exp \! \left[\, \int d^{4}x 
({\cal L}_{1}+{\cal L}_{\rm GF}) \right] 
\label{17}
\\
(&=& W'- \ln N_{1}\,) \,,
\nonumber 
\end{eqnarray}
%
where ${\frak D}{\cal M}'' \equiv
{\frak D}{\widetilde A}_{\mu}{}^{B} \, {\frak D}B_{\mu\nu} \, 
{\frak D}\varPhi\, {\frak D}h_{\mu\nu} \, {\frak D}c^{B} \, 
{\frak D}\bar{c}^{B} \,{\frak D}b^{B}$.

\section{Gauge fixing term}

We next define the gauge fixing term ${\cal L}_{\rm GF}$ in such a way that 
it satisfies the following three conditions:  
(i) ${\cal L}_{\rm GF}$ breaks the ^^ ^^ local" $SU(2)$ gauge 
invariance of ${\cal L}_{1}$.  
(ii) ${\cal L}_{\rm GF}$ remains invariant under the ^^ ^^ global" 
$SU(2)$ gauge transformation, provided that $c^{B}$, $\bar{c}^{B}$ 
and $b^{B}$ transform homogeneously under it, i.e., 
$\delta c^{B}=\epsilon^{BCD}c^{D} \lambda^{C}$, 
$\delta \bar{c}^{B}=\epsilon^{BCD} \bar{c}^{D} \lambda^{C}$, 
$\delta b^{B}=\epsilon^{BCD}b^{D} \lambda^{C}$, 
with $\lambda^{C}$ being ^^ ^^ constant" parameters.  
(iii) The quadratic part (or the asymptotic form in the limits  
$x^{0}\rightarrow\pm\infty$) of ${\cal L}_{\rm GF}$ 
has the same form as that of the Lorentz gauge-fixing term. 
The condition (i) is essential to quantize ${\widetilde A}_{\mu}{}^{B}$ 
(or $A_{\mu}{}^{B}$).  
The condition (ii) is necessary for preserving the global $SU(2)$ gauge 
invariance of ${\cal L}_{1}$ to guarantee conservation of the Noether current 
for the global $SU(2)$ gauge symmetry. 
It should be noted here that both of the local and global $SU(2)$ 
gauge transformations are converted into the corresponding local 
$U(1)$ gauge transformations via the transformation rule 
$gv(\phi_{0})=v(\phi'_{0})h(\phi_{0}, g)$. 
The local $SU(2)$ gauge invariance of ${\cal L}_{1}$ is then realized 
as invariance of ${\cal L}_{1}$ under the local $U(1)$ gauge transformation 
characterized by $h(\phi_{0}, g(x))$ with $g(x)$ being dependent on 
space-time coordinates $x^{\mu}$.  
The local $U(1)$ gauge transformation produced by the global $SU(2)$ gauge 
transformation is characterized by $h(\phi_{0}, g_{\rm c})$ with 
$g_{\rm c}$ being independent of $x^{\mu}$; 
the locality of this $U(1)$ gauge transformation is due to the 
$x^{\mu}$-dependence of $\phi_{0}{}^{i}$. 
The condition (iii) will be necessary for the proof of unitarity based on  
the Kugo-Ojima quartet mechanism \cite{Kugo-Ojima}. 
A gauge fixing term that satisfies the above conditions and  
that is appropriate for our discussion 
is given in terms of $\widetilde{A}_{\mu}{}^{B}$, 
$\tilde{e}_{\mu}{}^{B}\equiv e_{\mu}{}^{B}(\phi_{0})$, 
$\bar{\gamma}^{B}\equiv \bar{c}^{\,C} {\Bbb D}^{CB}(v(\phi_{0}))$ and 
$\beta^{B}\equiv b^{C} {\Bbb D}^{CB}(v(\phi_{0}))$ by 
\begin{eqnarray}
{\cal L}_{\rm GF} &=& {i\over{g_{0}^{2}}} \,
\boldsymbol\delta  \bigg[ \widetilde{D}^{\mu BC} \bar{\gamma}^{C} \!\cdot
( \widetilde{A}_{\mu}{}^{B}-\tilde{e}_{\mu}{}^{B} ) 
-{\alpha_{0}\over2}\,\bar{\gamma}^{B} \beta^{B} \bigg] \,,
\label{18}
\end{eqnarray}
%
where $\widetilde{D}_{\mu}{}^{BC}\equiv \delta^{BC}\partial_{\mu}
+\epsilon^{BC3}\widetilde{A}_{\mu}{}^{3}$, 
and $\alpha_{0}$ is a (bare) gauge parameter. 
The global $SU(2)$ gauge invariance of ${\cal L}_{\rm GF}$  
is realized as invariance of ${\cal L}_{\rm GF}$ under the local $U(1)$ gauge 
transformation characterized by $h(\phi_{0}, g_{\rm c})$. 
If $\widetilde{D}_{\mu}{}^{BC}$ in Eq. (\ref{18}) is replaced by 
$\widetilde{\nabla}_{\mu}{}^{BC}\equiv
\delta^{BC}\partial_{\mu}+\epsilon^{BCD} \tilde{e}_{\mu}{}^{D}$,  
then ${\cal L}_{\rm GF}$ reduces to the ordinary Lorenz gauge-fixing term for 
$A_{\mu}{}^{B}$, which preserves the global $SU(2)$ gauge invariance of 
${\cal L}_{1}$. Under the global $SU(2)$ gauge transformation, 
$\tilde{e}_{\mu}{}^{b}$ transform homogeneously in the same manner as 
$\widetilde{A}_{\mu}{}^{b}$, while $\tilde{e}_{\mu}{}^{3}$ transforms  
inhomogeneously in the same manner as $\widetilde{A}_{\mu}{}^{3}$. 
(In $\widetilde{\nabla}_{\mu}{}^{bc}$, $\tilde{e}_{\mu}{}^{3}$ plays the role 
of a gauge field as $\widetilde{A}_{\mu}{}^{3}$ plays in 
$\widetilde{D}_{\mu}{}^{bc}$.) 
From these facts we see that the condition (ii) is satisfied. 
Under the local $SU(2)$ gauge transformation, $\tilde{e}_{\mu}{}^{B}$ 
transform inhomogeneously in different manners than  
$\widetilde{A}_{\mu}{}^{B}$, so that the condition (i) is satisfied. 
Since $v(\phi_{0})$ approximates to  
$1+i\phi_{0}{}^{j}\delta_{j}{}^{B}T^{B}$ in the free limit, the condition 
(iii) is evidently satisfied.  (We can impose the condition $v(0)=1$ 
without loss of generality.)

The BRST transformation rules of 
$\widetilde{A}_{\mu}{}^{B}$,  $\tilde{e}_{\mu}{}^{B}$, 
$\gamma^{B}\equiv c^{C} {\Bbb D}^{CB}(v(\phi_{0}))$, $\bar{\gamma}^{B}$ and 
$\beta^{B}$ are determined from Eq. (\ref{8}) to be 
\begin{gather}
\begin{split}
\boldsymbol{\delta} \widetilde{A}_{\mu}{}^{b}
&=-\epsilon^{bc3}\widetilde{A}_{\mu}{}^{c} \tilde{c} \,,
\qquad 
\boldsymbol{\delta} \widetilde{A}_{\mu}{}^{3}
=\partial_{\mu}\tilde{c} \,, 
\\ 
\boldsymbol{\delta} \tilde{e}_{\mu}{}^{B} &= 
-\widetilde{\nabla}_{\mu}{}^{BC}(\gamma^{C}-\delta^{C3}\tilde{c}) \,, 
\\ 
\boldsymbol{\delta} \gamma^{B} &= 
-{1\over2}\epsilon^{BCD} \gamma^{C} \gamma^{D} 
+\epsilon^{BC3} \gamma^{C} \tilde{c} \,, 
\\ 
\boldsymbol{\delta} \bar{\gamma}^{B} &= i\beta^{B} 
-\epsilon^{BCD} \bar{\gamma}^{C} (\gamma^{D}-\delta^{D3} \tilde{c}) \,, 
\\ 
\boldsymbol{\delta} \beta^{B} &= 
\epsilon^{BCD} \beta^{C} (\gamma^{D}-\delta^{D3} \tilde{c}) \,,
\label{19}
\end{split}
\end{gather}
%
where $\tilde{c}\equiv c^{B}\Omega^{B}(\phi_{0})$. 
Carrying out the BRST transformation in the right hand side of 
Eq. (\ref{18}),  we obtain 
\begin{eqnarray}
{\cal L}_{\rm GF} &=& {1\over{g_{0}^{2}}} 
\bigg[ ( \beta^{b}+i\epsilon^{bDE} \gamma^{D} \bar{\gamma}^{E} )
\widetilde{D}^{\mu bc} \widetilde{A}_{\mu}{}^{c} 
\nonumber \\ 
& & -( \widetilde{D}^{\mu BC} \beta^{C}
+i\epsilon^{BCD} \bar{\gamma}^{C} \widetilde{D}^{\mu DE} \gamma^{E}  )
 ( \delta^{B3} \widetilde{A}_{\mu}{}^{3}
-\tilde{e}_{\mu}{}^{B} ) 
\nonumber \\ 
& &
-i\widetilde{D}^{\mu BC} \bar{\gamma}^{C} 
\widetilde{D}_{\mu}{}^{BD} \gamma^{D} 
+{\alpha_{0}\over2} \beta^{B}\beta^{B} \bigg] 
+ \mbox{total derivative} \,. 
\label{20}
\end{eqnarray}
%
The first term of Eq. (\ref{20}) vanishes owing to the equation 
\begin{eqnarray}
\widetilde{D}^{\mu bc} \widetilde{A}_{\mu}{}^{c}=
\widehat{D}^{\mu bc} \widehat{A}_{\mu}{}^{c}|_{\phi=\phi_{0}} =0 \,. 
\label{21}
\end{eqnarray}
%
Recall here that $\phi_{0}{}^{i}$ is a solution of Eq. (\ref{6}). 
Consequently ${\cal L}_{\rm GF}$ does not depend on the off-diagonal 
gluon fields $\widetilde{A}_{\mu}{}^{b}$, so that the effective action 
$W''$ can be written as  
\begin{eqnarray}
W''= \ln\int {\frak D}{\widetilde A}_{\mu}{}^{3} \, 
{\frak D}B_{\mu\nu}\, {\frak D}\varPhi\, {\frak D}h_{\mu\nu}\,
e^{( W''_{1}+W''_{\rm GF} )}
\,,
\label{22}
\end{eqnarray}
%
with 
\begin{eqnarray}
W''_{1}&=& \ln\int {\frak D}{\widetilde A}_{\mu}{}^{b} 
\exp \! \left[\, \int d^{4}x {\cal L}_{1} \right] , 
\label{23} 
\\ 
W''_{\rm GF}&=& \ln\int {\frak D}\gamma^{B} \,
{\frak D}\bar{\gamma}^{B} \,{\frak D}\beta^{B}
\exp \! \left[\, \int d^{4}x {\cal L}_{\rm GF} \right] .
\label{24}
\end{eqnarray}
%
Here the gauge invariance of the path-integral measures, i.e., 
${\frak D}c^{B}={\frak D}\gamma^{B}$, 
${\frak D}\bar{c}^{B}={\frak D}\bar{\gamma}^{B}$ and 
${\frak D}b^{B}={\frak D}\beta^{B}$, has been used.

\section{Effective Abelian gauge theory}

We now evaluate the path-integral over $\widetilde{A}_{\mu}{}^{b}$ 
in Eq. (\ref{23}), rewriting Eq. (\ref{13}) in the form 
$ {\cal L}_{1}={\cal L}_{1}^{(0)}+{1\over2} \widetilde{A}_{\mu}{}^{b} 
{\cal H}_{1}^{\mu b,\nu c} \widetilde{A}_{\nu}{}^{c} +O_{1}(\varLambda^{-2})$ 
up to total derivatives. Here ${\cal L}_{1}^{(0)}$ consists of the terms 
that contain no $\widetilde{A}_{\mu}{}^{b}$, i.e., 
\begin{align}
{\cal L}_{1}^{(0)} = \: & \frac{\varLambda^{4}}{32\pi^{2}}
-{1\over4}
\bigg(\frac{1}{g_{0}^{2}}-\frac{1}{24\pi^{2}} \ln \frac{\mu}{\varLambda}
-\frac{\kappa^{2}}{q_{1}} \bigg) 
{\cal F}_{\mu\nu}{\cal F}^{\mu\nu} 
\nonumber \\ 
& +{i\over2} \kappa B_{\mu\nu}{\cal F}^{\mu\nu}
-\frac{q_{1}}{4} B_{\mu\nu}B^{\mu\nu} 
-\frac{q_{2}}{4} \,\varPhi^{2} 
-\frac{q_{3}}{4} h_{\mu\nu}h^{\mu\nu} \,.
\label{25}
\end{align}
%
Recall the fact that $O_{1}(\varLambda^{-2})$ has been found at the one-loop 
level in evaluating the path-integral over $\phi^{i}$. 
Then we see that the terms in 
$O_{1}(\varLambda^{-2})$ which contain ${\widetilde A}_{\mu}{}^{b}$ 
contribute as two-loop effects to the one-loop effective action obtained 
by the integration over ${\widetilde A}_{\mu}{}^{b}$ in Eq. (\ref{23}), 
provided that the one-loop approximation is carried out around the classical 
solution ${\widetilde A}_{\mu}{}^{b}=0$. 
Because the two-loop effects are now irrelevant to our discussion, 
it is sufficient for evaluating the path-integral over 
${\widetilde A}_{\mu}{}^{b}$ to consider 
${1\over2} \widetilde{A}_{\mu}{}^{b} 
{\cal H}_{1}^{\mu b,\nu c} \widetilde{A}_{\nu}{}^{c}$ only. 
After the use of the commutation relation 
$[\widetilde{D}_{\mu}, \widetilde{D}_{\nu}]^{bc}\widetilde{A}_{\rho}{}^{c}=
\epsilon^{3bc}{\cal F}_{\mu\nu}\widetilde{A}_{\rho}{}^{c}$ and Eq. (\ref{21}), 
${\cal H}_{1}^{\mu b,\nu c}$ becomes 
the Laplace-type operator 
\begin{eqnarray}
{\cal H}_{1}^{\mu b,\nu c}  
&=& \frac{1}{g_{0}^{2}} \eta^{\mu\nu} 
\widetilde{D}_{\rho}{}^{bd} \widetilde{D}^{\rho dc} 
+\eta^{\mu\nu}\delta^{bc} 
\bigg( \frac{m_{0}^{2}}{g_{0}^{2}}-\frac{\varLambda^{2}}{16\pi^{2}} 
+i \varPhi \bigg) 
+i\delta^{bc} h^{\mu\nu} 
\nonumber \\ 
& & +\epsilon^{bc3} \bigg\{ \bigg( \frac{2}{g_{0}^{2}}
-\frac{\kappa}{q_{1}}\bigg)
{\cal F}^{\mu\nu}-iB^{\mu\nu} \bigg\} \,.
\label{26}
\end{eqnarray} 
%
Supposing  
$m_{0}/g_{0}\geq \varLambda/4\pi$, 
we carry out the Gaussian integration over $\widetilde{A}_{\mu}{}^{b}$ in  
Eq. (\ref{23}) to get an expression written with the aid of 
a proper-time $\tau\,$: 
\begin{align}
\int d^{4}x {\cal L}_{1}^{(1)} 
& = \ln\int {\frak D}{\widetilde A}_{\mu}{}^{b} 
\exp \! \left[\,  {1\over2} \int d^{4}x
\widetilde{A}_{\mu}{}^{b} {\cal H}_{1}^{\mu b,\nu c} \widetilde{A}_{\nu}{}^{c} 
\right] \! +\ln N_{2}  
\nonumber \\ 
&  =\int d^{4}x {1\over2} \sum_{\mu, b} 
\int_{1/\varLambda^{2}}^{\infty} \frac{d\tau}{\tau} 
\langle\!\langle x, {}_{\mu}{}^{b}| e^{-\tau \widehat{\cal H}_{1}} 
|x, {}^{\mu b} \rangle\!\rangle \,. 
\label{27} 
\end{align}
%
(Later on we will point out that $m_{0}/g_{0}=\varLambda/4\pi$ is 
a reasonable condition, see Eq. (\ref{60}).) 
The ket vectors $|x,{}^{\mu b} \rangle\!\rangle $ are defined by 
$|x,{}^{\mu b} \rangle\!\rangle 
=\sum_{C}|x,{}^{\mu C} \rangle {\Bbb D}^{Cb}(v(\phi_{0}))$ 
with the basis vectors $|x,{}^{\mu B} \rangle $ that satisfy 
$\langle x,{}_{\mu}{}^{B} |y,{}^{\nu C} \rangle
=\delta^{4}(x-y)\delta_{\mu}{}^{\nu} \delta^{BC}$ 
and that transform homogeneously under the $SU(2)$ gauge 
transformations, 
i.e., $\delta |x,{}^{\mu B} \rangle 
=\epsilon^{BCD} |x,{}^{\mu D} \rangle \lambda^{C}$. 
Then $|x,{}^{\mu b} \rangle\!\rangle $ transform in the same manner as 
$\widetilde{A}_{\mu}{}^{b}$ (see Eq. (\ref{10a})), 
so that the gauge invariance 
of the operator 
\begin{eqnarray}
\widehat{\cal H}_{1} \equiv 
\int d^{4}x \sum_{\mu,b}\sum_{\nu,c} 
|x, {}^{\mu b} \rangle\!\rangle {\cal H}_{1 \mu}{}^{b,\nu c}(x)
\langle\!\langle x, {}_{\nu}{}^{c}| 
\label{28} 
\end{eqnarray}
%
is guaranteed. In this way the gauge invariance of ${\cal L}_{1}^{(1)}$  
is confirmed.  
Using the heat kernel equation for the matrix elements 
$\langle\!\langle x, {}_{\mu}{}^{b} | e^{-\tau \widehat{\cal H}_{1}} 
|y, {}^{\nu c} \rangle\!\rangle$, 
we can approximately calculate the effective 
action in Eq. (\ref{27}). In the process of calculation, 
we have to take account of the Dirac 
string singularity occurring in 
$[\partial_{\mu},\partial_{\nu}]{\Bbb D}^{BC}(v^{-1}(\phi_{0}))
=-\epsilon^{3BD} \varSigma_{\mu\nu}(\phi_{0}){\Bbb D}^{DC}(v^{-1}(\phi_{0}))$. 
The result of the calculation reads 
\begin{align}
{\cal L}_{1}^{(1)}=\:& \frac{\varLambda^{4}}{8\pi^{2}}
-\frac{g_{0}^{2}\varLambda^{2}}{4\pi^{2}}i\varPhi 
-{1\over{16\pi^{2}}} \ln \frac{\mu}{\varLambda}
\bigg[ \bigg\{ 4\bigg(1-\frac{\kappa g_{0}^{2}}{2q_{1}}\bigg)^{2} 
-{2\over3} \bigg\} {\cal F}_{\mu\nu}{\cal F}^{\mu\nu} 
\nonumber \\ 
& -4ig_{0}^{2}\bigg(1-\frac{\kappa g_{0}^{2}}{2q_{1}}\bigg) 
B_{\mu\nu}{\cal F}^{\mu\nu} -g_{0}^{4}B_{\mu\nu}B^{\mu\nu} 
-4g_{0}^{4}\varPhi^{2} -g_{0}^{4}h_{\mu\nu}h^{\mu\nu} \bigg] 
\nonumber \\ 
& +\frac{g_{0}^{4}}{192\pi^{2}\varLambda^{2}} 
\bigg( \partial_{\mu}B_{\nu\rho} \partial^{\mu}B^{\nu\rho} 
+4\partial_{\mu}\varPhi \partial^{\mu}\varPhi 
+\partial_{\mu}h_{\nu\rho} \partial^{\mu}h^{\nu\rho} \bigg) 
\nonumber \\ 
& + O_{2\rm h}(\varLambda^{-2}) 
\label{29}
\end{align}
%
after making the replacement 
$\varPhi\rightarrow\varPhi+i(m_{0}^{2}/g_{0}^{2}-\varLambda^{2}/16\pi^{2})$. 
Here $O_{2\rm h}(\varLambda^{-2})$ consists of terms of higher order 
in ${\cal F}_{\mu\nu}$, $B_{\mu\nu}$, $\varPhi$, $h_{\mu\nu}$ and/or 
their derivatives, including such higher derivative terms as 
$\partial_{\mu}{\cal F}_{\nu\rho} \partial^{\mu}{\cal F}^{\nu\rho}$ and 
$\partial_{\mu}B_{\nu\rho} \partial^{\mu}{\cal F}^{\nu\rho}$. 
It should be stressed that the kinetic terms of $B_{\mu\nu}$, $\varPhi$ 
and $h_{\mu\nu}$ have been induced by virtue of quantum effect of 
$\widetilde{A}_{\mu}{}^{b}$. 
In this sense $B_{\mu\nu}$, $\varPhi$ and $h_{\mu\nu}$ are no longer 
auxiliary fields. 
(As for $B_{\mu\nu}$, such an induced kinetic term has been found 
in various contexts \cite{Ellwanger,Kondo 2,Freire,Kondo 3}.)

Next we consider the path-integral over $\gamma^{B}$, $\bar{\gamma}^{B}$ 
and $\beta^{B}$ in Eq. (\ref{24}). 
In the case $\alpha_{0}\neq 0$, the integration over $\beta^{B}$ becomes 
a simple Gaussian integration and can readily be done to get 
$\int d^{4}x {\cal L}'_{\rm GF}=\ln \int {\frak D}\beta^{B}
\exp \! \big[ \int d^{4}x {\cal L}_{\rm GF} \big]+\ln N_{3}\,$. 
In order to perform the integration over 
$\gamma^{B}$ and $\bar{\gamma}^{B}$, 
we rewrite ${\cal L}'_{\rm GF}$ in the form 
$ {\cal L}'_{\rm GF}={\cal L}_{\rm GF}^{(0)}
+i \bar{\gamma}^{B} {\cal H}_{\rm GF}^{BC} \gamma^{C}$  
up to total derivatives. Here  
${\cal L}_{\rm GF}^{(0)}$ and ${\cal H}_{\rm GF}^{BC}$ are given by 
\begin{align} 
{\cal L}_{\rm GF}^{(0)}
=&-\frac{1}{2\alpha_{0} g_{0}^{2}} (
\partial^{\mu} a_{\mu} \partial^{\nu} a_{\nu} 
+ \widetilde{\nabla}^{\mu bc} \tilde{e}_{\mu}{}^{c} 
\widetilde{\nabla}^{\nu bd} \tilde{e}_{\nu}{}^{d} 
-2\epsilon^{3bc}a_{\mu} \tilde{e}^{\mu b} 
\widetilde{\nabla}^{\nu cd} \tilde{e}_{\nu}{}^{d} 
\nonumber \\ 
&
+a_{\mu}a_{\nu} \tilde{e}^{\mu b} \tilde{e}^{\nu b} ) \,,
\label{30}
\\
{\cal H}_{\rm GF}^{BC}=\:& \frac{1}{g_{0}^{2}} \bigg[ \,
\widetilde{\varDelta}_{\mu}{}^{BD} \widetilde{\varDelta}^{\mu DC} 
-{1\over2}\epsilon^{BCa} \widetilde{D}^{\mu ad} \tilde{e}_{\mu}{}^{d} 
+{1\over2}\epsilon^{BC3} \partial^{\mu}a_{\mu} 
\nonumber \\ 
&
+{1\over4} \{ \delta^{BC} 
( a_{\mu}a^{\mu} +\tilde{e}_{\mu}{}^{d} \tilde{e}^{\mu d} ) 
-(\delta^{B3} \widetilde{A}_{\mu}{}^{3} -\tilde{e}_{\mu}{}^{B} )
(\delta^{C3} \widetilde{A}^{\mu 3} -\tilde{e}^{\mu C} ) \} 
\bigg] 
\label{31}
\end{align} 
%
with  $a_{\mu}\equiv\widetilde{A}_{\mu}{}^{3}-\tilde{e}_{\mu}{}^{3}$, 
$\,\widetilde{\nabla}_{\mu}{}^{bc} 
\equiv \delta^{bc}\partial_{\mu}+\epsilon^{bc3}\tilde{e}_{\mu}{}^{3}$ 
and 
\begin{align}
\widetilde{\varDelta}_{\mu}{}^{BC} \equiv \:& \delta^{BC}\partial_{\mu}
+{1\over2}\epsilon^{BCd} \tilde{e}_{\mu}{}^{d} 
+{1\over2}\epsilon^{BC3}
(\widetilde{A}_{\mu}{}^{3}+ \tilde{e}_{\mu}{}^{3} ) \,.
\label{32}
\end{align}
%
Note that under the global $SU(2)$ gauge  
transformation, $a_{\mu}$ is invariant, while 
${1\over2}(\widetilde{A}_{\mu}{}^{3}+ \tilde{e}_{\mu}{}^{3})$ 
transforms inhomogeneously in the same manner as $\widetilde{A}_{\mu}{}^{3}$.  
In $\widetilde{\varDelta}_{\mu}{}^{bc}$, 
${1\over2}(\widetilde{A}_{\mu}{}^{3}+ \tilde{e}_{\mu}{}^{3})$  
plays the role of a gauge field. 
The Gaussian integration over $\gamma^{B}$ and 
$\bar{\gamma}^{B}$ is carried out to get 
\begin{align}
\int d^{4}x {\cal L}_{\rm GF}^{(1)} 
&  = \ln\int {\frak D}\gamma^{B}\, {\frak D}\bar{\gamma}^{B} 
\exp \! \left[\,  i \int d^{4}x
\bar{\gamma}^{B} {\cal H}_{\rm GF}^{BC} \gamma^{C} 
\right] \! +\ln N_{4}  
\nonumber \\ 
&  =-\int d^{4}x  \sum_{B} 
\int_{1/\varLambda^{2}}^{\infty} \frac{d\tau}{\tau} 
\langle\!\langle x, {}^{B} | e^{-\tau \widehat{\cal H}_{\rm GF}} 
|x, {}^{B} \rangle\!\rangle \,,
\label{33} 
\end{align}
%
where the ket vectors $|x, {}^{B} \rangle\!\rangle $ are defined by 
$|x, {}^{B} \rangle\!\rangle 
=\sum_{C}|x, {}^{C} \rangle {\Bbb D}^{CB}(v(\phi_{0}))$ 
with the orthonormal basis vectors $|x, {}^{B} \rangle $ 
that transform homogeneously under the $SU(2)$ gauge transformations. 
Then $|x, {}^{B} \rangle\!\rangle $ obey the gauge transformation rule  
same as that of $\gamma^{B}$ and $\bar{\gamma}^{B}$, so that the operator 
$\widehat{\cal H}_{\rm GF} \equiv 
\int d^{4}x \sum_{B,C} 
|x, {}^{B} \rangle\!\rangle {\cal H}_{\rm GF}^{BC}(x) 
\langle\!\langle x, {}^{C}| $ 
is gauge invariant. 
Consequently the gauge invariance of ${\cal L}_{\rm GF}^{(1)}$ is also 
confirmed.  
Applying the heat kernel method to calculating the $\tau$-integral in  
Eq. (\ref{33}), we obtain 
\begin{align}
{\cal L}_{\rm GF}^{(1)}=\,& -\frac{3\varLambda^{4}}{32\pi^{2}}
+\frac{\varLambda^{2}}{32\pi^{2}} 
(a_{\mu}a^{\mu} + \tilde{e}_{\mu}{}^{b}\tilde{e}^{\mu b}) 
\nonumber \\ 
&+{1\over 8\pi^{2}} \ln \frac{\mu}{\varLambda}
\bigg\{ -{1\over24} ({\cal F}_{\mu\nu} {\cal F}^{\mu\nu} 
+{\cal F}_{\mu\nu} J^{\mu\nu} + J_{\mu\nu}J^{\mu\nu}) 
\nonumber \\ 
& -{1\over4}\partial^{\mu} a_{\mu} \partial^{\nu} a_{\nu} 
+{1\over16}(a_{\mu}a^{\mu})^{2}
+{1\over2} \epsilon^{3bc}a_{\mu} \tilde{e}^{\mu b} 
\widetilde{\nabla}^{\nu cd} \tilde{e}_{\nu}{}^{d} 
\nonumber \\ 
& +{1\over24}a_{\mu}a^{\mu}\tilde{e}_{\nu}{}^{b}\tilde{e}^{\nu b}
-{1\over6}a_{\mu}a_{\nu}\tilde{e}^{\mu b}\tilde{e}^{\nu b}
-{1\over4} \widetilde{\nabla}^{\mu bc} \tilde{e}_{\mu}{}^{c} 
\widetilde{\nabla}^{\nu bd} \tilde{e}_{\nu}{}^{d} 
\nonumber \\ 
& 
+{1\over16}(\tilde{e}_{\mu}{}^{b}\tilde{e}^{\mu b})^{2} \bigg\} 
+O_{3}(\varLambda^{-2}) 
\label{34}
\end{align}
%
with  
$J_{\mu\nu}=J_{\mu\nu}(\phi_{0})
\equiv\epsilon^{3bc}\tilde{e}_{\mu}{}^{b} \tilde{e}_{\nu}{}^{c}$. 
By using the Maurer-Cartan formula in Eq. (\ref{4}) 
at $\phi^{i}=\phi_{0}{}^{i}$, ${\cal F}_{\mu\nu}$ can be written as 
\begin{eqnarray}
{\cal F}_{\mu\nu}=f_{\mu\nu}+J_{\mu\nu} \,, 
\label{35}
\end{eqnarray}
%
with $f_{\mu\nu}\equiv \partial_{\mu}a_{\nu}-\partial_{\nu}a_{\mu}$.

The exponent in Eq. (\ref{22}) is thus found to be 
$W''_{1}+W''_{\rm GF}=\int d^{4}x {\cal L}_{\rm tot}-\ln(N_{2}N_{3}N_{4})$  
with 
\begin{align}
{\cal L}_{\rm tot}
= \:&{\cal L}_{1}^{(0)}+{\cal L}_{1}^{(1)}
+{\cal L}_{\rm GF}^{(0)}+{\cal L}_{\rm GF}^{(1)} 
+O_{1}'(\varLambda^{-2}) \,, 
\label{36}
\end{align}
%
where $O_{1}'(\varLambda^{-2})$ denotes a part of $O_{1}(\varLambda^{-2})$ 
that consists of the terms containing ${\cal F}_{\mu\nu}$ but no  
$\widetilde{A}_{\mu}{}^{b}$. 
In what follows, we are concerned with 
the effective Abelian gauge theory (EAGT) defined by 
the Lagrangian ${\cal L}_{\rm tot}$. 
As has been expected, ${\cal L}_{\rm tot}$ remains invariant 
under the $U(1)$ gauge transformation characterized by 
$h(v(\phi_{0}), g_{\rm c})$. 
Since this transformation is produced by the global $SU(2)$ 
gauge transformation due to the action of $g_{\rm c}$, 
we should consider that the global $SU(2)$ gauge symmetry 
is maintained in the EAGT and 
is realized in ${\cal L}_{\rm tot}$ in a nonlinear way. 
In fact, it is possible to derive the non-Abelian Noether current 
for the global $SU(2)$ gauge invariance of ${\cal L}_{\rm tot}$. 
This current is written with the use of the $H$-compensators 
$\varOmega^{B}(\phi_{0})$.

\section{Static potentials}

Considering ${\cal L}_{\rm tot}$ as a classical Lagrangian, let us 
investigate what kind of potential arises from propagation 
of $B_{\mu\nu}$ at the tree level of the EAGT. 
With ${\cal L}_{\rm tot}$ in hand, the bare propagator of $B_{\mu\nu}$ 
follows from the kinetic term and  mass term of $B_{\mu\nu}$ together with 
a source term of $B_{\mu\nu}$. 
%
In addition to the linear terms in $B_{\mu\nu}$ explicitly written in  
Eq. (\ref{25}) and Eq. (\ref{29}), further linear terms 
in $B_{\mu\nu}$ exist in $O_{2\rm h}(\varLambda^{-2})$. 
Fortunately, if $\bar{\kappa}\equiv 2q_{1}/g_{0}^{2}$ 
is chosen to be the constant $\kappa$ in ${\cal L}_{\rm tot}$, 
these linear terms 
as well as the $B_{\mu\nu} {\cal F}^{\mu\nu}$ term in Eq. (\ref{29}) vanish 
in ${\cal L}_{1}^{(1)}$ and it becomes easier to make a discussion. 
On putting $\kappa=\bar{\kappa}$, the terms relevant to deriving 
the bare propagator of $B_{\mu\nu}$ are displayed, after being rescaled 
$B_{\mu\nu}$ as 
$B_{\mu\nu} \rightarrow 4\sqrt{3}\pi\varLambda g_{0}^{-2} B_{\mu\nu}$, 
as follows: 
\begin{eqnarray}
{\cal L}_{B}^{(0)} &\equiv&
-{1\over4} B_{\mu\nu}( \square +M_{1}^{2}) B^{\mu\nu}
+{i\over2} M_{2} B_{\mu\nu} {\cal F}^{\mu\nu} \,,
\label{37}
\end{eqnarray}
%
where $\square\equiv\partial_{\mu}\partial^{\mu}
=-\partial_{\mu}\partial_{\mu}$ and 
\begin{eqnarray}
M_{1}& \equiv & \frac{4\sqrt{3}\pi}{g_{0}^{2}} 
\bigg( q_{1}-\frac{g_{0}^{4}}{4\pi^{2}} \ln\frac{\mu}{\varLambda} \bigg)^{1/2}
\varLambda  \,, 
\label{38}
\\ 
M_{2}& \equiv & \frac{8\sqrt{3}\pi q_{1}}{g_{0}^{4}} \varLambda \,.
\label{39}
\end{eqnarray}
%
Since $J^{\mu\nu}$ consists only of the classical fields $\phi_{0}{}^{i}$, 
we can treat $J^{\mu\nu}$ as a classical antisymmetric 
tensor current. 
Then we see that ${\cal L}_{B}^{(0)}$ is very analogous to a part of the 
covariantly gauge fixed version of the Lagrangian that defines a massive 
Abelian antisymmetric tensor gauge theory (MAATGT) with antisymmetric 
tensor current (ATC) \cite{Deguchi-Kokubo}.\footnote{
\setlength{\baselineskip}{6mm}
If the vorticity tensor 
current is chosen as the ATC, 
the MAATGT with ATC reduces to a dual theory of the extended dual 
Abelian Higgs model (EDAHM) in the London limit 
\cite{Antonov-Ebert,Deguchi-Kokubo}. }
Obviously ${\cal L}_{B}^{(0)}$ describes a massive rank-2 antisymmetric 
tensor field coupled with $f^{\mu\nu}$ and $J^{\mu\nu}$.  
For any mass scale characterized by $\mu (\,\leq \! \varLambda)$,  
positivity of $M_{1}^{2}$ is guaranteed as long as appropriate constants are 
chosen to be $k_{1}$ and $k_{2}$. Let us denote by ${\cal L}_{B}^{(1)}$ 
a part of $O_{2\rm h}(\varLambda^{-2})$ that consists of the terms 
containing $B_{\mu\nu}$. Noting the fact that the source term 
${1\over2}i M_{2} B_{\mu\nu} J^{\mu\nu}$ in ${\cal L}_{B}^{(0)}$ exists 
in the exponent of Eq. (\ref{22}), 
we represent $B_{\mu\nu}$ in ${\cal L}_{B}^{(1)}$ as 
the functional derivative $-iM_{2}^{-1}\delta/\delta J^{\mu\nu}$. 
Thereby the integration over $B_{\mu\nu}$ in Eq. (\ref{22}) reduces to 
a Gaussian integration, which can be carried out to obtain 
\begin{eqnarray}
W''= \ln\int {\frak D}{\widetilde A}_{\mu}{}^{3} \, {\frak D}\varPhi\, 
{\frak D}h_{\mu\nu} \exp \! \left[\,\int d^{4}x {\cal L}'_{\rm tot} 
\right] \,,
\label{40}
\end{eqnarray}
where ${\cal L}'_{\rm tot}$ is given by 
\begin{align}
\int d^{4}x {\cal L}'_{\rm tot} = & \: W''_{1}+W''_{\rm GF} 
-\int d^{4}x \big({\cal L}_{B}^{(0)}+{\cal L}_{B}^{(1)}\big) +\ln Z_{B} 
\label{41}
\end{align}
%
with 
\begin{align}
Z_{B}\equiv& \:{1\over N_{5}} \exp \! \left[\, \int d^{4}x 
{\cal L}_{B}^{(1)} | 
\raisebox{-0.65ex}{\mbox{\scriptsize{$B_{\mu\nu}= 
-iM_{2}^{-1} \dfrac{\delta}{\delta J^{\mu\nu}}$ }}} 
\right] 
\nonumber \\ 
& \times 
\exp \! \left[\, \int d^{4}x 
\bigg( -{1\over4}{\cal F}_{\mu\nu}
\frac{M_{2}^{2}}{\square+M_{1}^{2}}\,{\cal F}^{\mu\nu} 
\bigg) \right] 
\nonumber \\ 
=& \: {1\over N_{5}} \exp \! \left[\, \int d^{4}x 
\bigg( -{1\over4}{\cal F}_{\mu\nu}
\frac{M_{2}^{2}}{\square+M_{1}^{2}}\,{\cal F}^{\mu\nu} 
+\cdots \bigg) \right] \,.   
\label{42}
\end{align}
%
The bare propagator of $B_{\mu\nu}$ has 
been found in the exponent of Eq. (\ref{42}). 
The remaining terms indicated by the dots may be calculated, 
at least formally, by a perturbative method.

Together with deriving the potential due to propagation of $B_{\mu\nu}$, 
we also try to derive the potential due to propagation of $a_{\mu}$, 
rather than ${\widetilde A}_{\mu}{}^{3}$, at the tree level. 
To this end, we pay attention to 
a part of ${\cal L}'_{\rm tot}$ which consists of a relevant source term 
for $a_{\mu}$ and of the terms that are quadratic in $a_{\mu}$  
or its first order derivatives and contain no other fields. 
Here, as a relevant term to our discussion, we also take the local term 
$-{1\over4} (M_{2}/M_{1})^{2} (f_{\mu\nu}+2J_{\mu\nu})f^{\mu\nu}$ 
that is given as a leading term of the series expansion 
$(\square+M_{1}^{2})^{-1}=M_{1}^{-2}(1-M_{1}^{-2}\square+\cdots)$ 
in the non-local term explicitly written in the exponent of Eq. (\ref{42}). 
The part of ${\cal L}'_{\rm tot}$ that we are 
concerned now with is thus found from Eqs. (\ref{25}), 
(\ref{29}), (\ref{30}), (\ref{34}) and (\ref{42}) to be 
\begin{align}
{\cal L}_{a}^{(0)}\equiv& -\frac{\rho_{1}}{4g_{0}^{2}} f_{\mu\nu} f^{\mu\nu} 
-\frac{\rho_{2}}{2\alpha_{0} g_{0}^{2}}\,
\partial^{\mu} a_{\mu} \partial^{\nu} a_{\nu} 
+\frac{\varLambda^{2}}{32\pi^{2}}\, a_{\mu} a^{\mu} 
-\frac{\rho_{3}}{2g_{0}^{2}} f_{\mu\nu} J^{\mu\nu} 
\label{43}
\end{align}
%
with  
\begin{eqnarray}
\rho_{1}&\equiv& 1+\frac{13g_{0}^{2}}{16\pi^{2}} \ln\frac{\mu}{\varLambda} \,,
\label{44} 
\\ 
\rho_{2}&\equiv& 
1+\frac{\alpha_{0} g_{0}^{2}}{16\pi^{2}} \ln\frac{\mu}{\varLambda} \,, 
\label{45} 
\\ 
\rho_{3}&\equiv& 1+\frac{79g_{0}^{2}}{96\pi^{2}} \ln\frac{\mu}{\varLambda} \,.
\label{46}
\end{eqnarray}
%
Here $(M_{2}/M_{1})^{2}$ has been approximated by using the approximate 
formula $(1+O(\hbar))^{n}\approx 1+nO(\hbar)$. 
The renormalized coupling constant, $g(\mu)$, is read from the first 
term of Eq. (\ref{43}) to be $g(\mu)=g_{0}/\sqrt{\rho_{1}}$. 
Since $g(\mu)$ decreases with $\mu$, 
the EAGT enjoys the property of asymptotic freedom. 
(Because of the presence of the NG scalar fields $\phi^{i}$, 
behavior of $g(\mu)$ is somewhat different 
from that of the running coupling constant of $SU(2)$ YM theory 
\cite{Kondo 1,Reinhardt}.) 
The last term of Eq. (\ref{43}) is certainly a source term of 
$a_{\mu}$, as can be seen by rewriting it as 
$-{1\over2}\rho_{3}g_{0}^{-2}f_{\mu\nu}J^{\mu\nu}
=\rho_{3}g_{0}^{-2}a_{\mu}j^{\mu}$ 
up to total derivatives. Here $j^{\mu}$ is defined by 
$j^{\mu}=\partial_{\nu}J^{\nu\mu}$. 
By virtue of the antisymmetric property of $J^{\mu\nu}$, i.e., 
$J^{\mu\nu}=-J^{\nu\mu}$, 
$j^{\mu}$ satisfies the conservation law $\partial_{\mu}j^{\mu}=0$. 
In accordance with the treatment of $J^{\mu\nu}$, we treat $j^{\mu}$   
as a classical vector current. 
Since $j^{\mu}$ is a source of $a_{\mu}$, $j^{\mu}$ will be 
a color-electric current. 
Let us denote by ${\cal L}_{a}^{(1)}$ a part of ${\cal L}'_{\rm tot}$ that 
consists of the terms containing $a_{\mu}$ and that 
includes no the terms already included in ${\cal L}_{a}^{(0)}$. 
(${\cal L}_{a}^{(1)}$ includes the higher derivative terms 
$-{1\over4} (M_{2}/M_{1})^{2}(f_{\mu\nu}+2J_{\mu\nu})\square^{n} 
f^{\mu\nu}$ $(n=1,2,\dots)$ 
arising in the series expansion in the exponent of Eq. (\ref{42}).) 
Utilizing the source term $\rho_{3}g_{0}^{-2}a_{\mu}j^{\mu}$ 
that occurs in the exponent of Eq. (\ref{40}) through Eq. (\ref{43}), 
we represent $a_{\mu}$ in ${\cal L}_{a}^{(1)}$ as the functional derivative  
$\rho_{3}^{-1}g_{0}^{2} \delta/\delta j^{\mu}$. 
Then, noticing the invariance of path-integral measure 
${\frak D}{\widetilde A}_{\mu}{}^{3}={\frak D}a_{\mu}$, we carry out 
the integration over $a_{\mu}$ in Eq. (\ref{40}) to obtain  
\begin{eqnarray}
W''&=& \ln\int {\frak D}\varPhi\, 
{\frak D}h_{\mu\nu} 
\exp \! \left[\,\int d^{4}x 
\big( {\cal L}'_{\rm tot}-{\cal L}_{a}^{(0)}-{\cal L}_{a}^{(1)} \big)
\right] \! Z_{a} 
\label{47}
\end{eqnarray}
with  
\begin{align}
Z_{a}\equiv& \:{1\over N_{6}} \exp \! \left[\, \int d^{4}x 
{\cal L}_{a}^{(1)} |
\raisebox{-0.57ex}{\mbox{\scriptsize{$a_{\mu}= 
\dfrac{g_{0}^{2}}{\rho_{3}} \dfrac{\delta}{\delta j^{\mu}}$ }}} 
\right] 
\nonumber \\ 
& \times \exp \! \left[\, \int d^{4}x 
\bigg( -{1\over2} j_{\mu}
\frac{(\rho_{3}/g_{0})^{2}}{\rho_{1}\square
+g_{0}^{2}\varLambda^{2}/16\pi^{2}} \,
j^{\mu} \bigg) \right] \,.   
\label{48}
\end{align}
%
In deriving Eq. (\ref{47}), the conservation law of $j^{\mu}$ has 
been used, so that $Z_{a}$ turns out to be independent of the renormalized 
gauge parameter $\alpha_{0}/\rho_{2}$.

At the present stage, the effective action $W''$ takes the following form: 
\begin{eqnarray}
W''&=& W_{j}^{(0)}+W_{J}^{(0)}+\int d^{4}x {\cal L}_{2} 
\nonumber \\ 
& & + \ln\int {\frak D}\varPhi\, {\frak D}h_{\mu\nu}
\exp \! \left[\,\int d^{4}x {\cal L}_{3} \right] 
-\ln \prod_{l=1}^{6}N_{l} 
\label{49} 
\end{eqnarray}
%
with 
\begin{align}
W_{j}^{(0)}=& \int d^{4}x \bigg( -{1\over2}\, j_{\mu}
\frac{(\rho_{3}/g_{0})^{2}}{\rho_{1}\square 
+g_{0}^{2}\varLambda^{2}/16\pi^{2}} \,j^{\mu} \bigg) \,, 
\label{50} 
\\ 
W_{J}^{(0)}=& \int d^{4}x \bigg( -{1\over4} J_{\mu\nu}
\frac{M_{2}^{2}}{\square+M_{1}^{2}}\,J^{\mu\nu} \bigg) \,. 
\label{51}
\end{align}
%
Here ${\cal L}_{2}$ consists of the terms containing $\phi_{0}{}^{i}$ alone, 
some of which terms depend on $\phi_{0}{}^{i}$ 
through $J^{\mu\nu}$ or $j^{\mu}$, while 
${\cal L}_{3}$ consists of the remaining terms containing 
$\varPhi$ and/or $h_{\mu\nu}$. 
The functional $W_{J}^{(0)}$ is nothing but a part of the term explicitly  
written in the exponent of Eq. (\ref{42}). 
It is now obvious that the effective action $W''$ is a functional of 
the classical fields $\phi_{0}{}^{i}$ alone. 
Now, we would like to point out that the terms analogous to $W_{j}^{(0)}$ and 
$W_{J}^{(0)}$ have been found in a form of the generating functional 
characterizing the MAATGT with ATC \cite{Deguchi-Kokubo}. 
In this theory, a composite of the Yukawa and the linear potentials 
was obtained from the generating functional.  
It is hence expected that similar potential is derived 
from $W_{j}^{(0)}+W_{J}^{(0)}$.

To confirm this, we follow a simple procedure discussed 
in Ref. \cite{Deguchi-Kokubo}:  
Recalling $j^{\mu}=\partial_{\nu}J^{\nu\mu}$, 
we rewrite it in the integral form 
\begin{eqnarray}
J^{\mu\nu}=\frac{1}{n\cdot\partial}(n^{\mu}j^{\nu}-n^{\nu}j^{\mu}) 
\label{52}
\end{eqnarray}
%
with the aid of a constant vector $n^{\mu}$, 
where $n\cdot\partial\equiv n^{\mu}\partial_{\mu}$. 
(The vector $n^{\mu}$ may be understood to be a set of integration  
constants.) Substituting Eq. (\ref{52}) into Eq. (\ref{51}), we have 
\begin{eqnarray}
W_{J}^{(0)}=W_{J}^{(0)}[j^{\mu}]
&\equiv& 
\int d^{4}x \bigg[\,
{1\over2}\,j_{\mu} \bigg\{ 
\frac{M_{2}^{2}}{\square+M_{1}^{2}} 
\frac{n^{2}}{(n\cdot\partial)^{2}}
\bigg({\delta^{\mu}}_{\nu} -\frac{n^{\mu}n_{\nu}}{n^{2}} \bigg)
\bigg\} j^{\nu} \bigg] \,, 
\label{53}
\end{eqnarray}
%
where $n^{2}\equiv n_{\mu}n^{\mu}$. 
As a result, $W_{J}^{(0)}$ is expressed as a functional of the vector 
current $j^{\mu}$.

In order to evaluate the static potential based on 
$W_{j}^{(0)}+W_{J}^{(0)}[j^{\mu}]$, we now replace $j^{\mu}$ with the static 
current $j^{\mu}_{Q}(x)\equiv{\delta^{\mu}}_{4} Q \left\{ \delta^{3}({\bold x}
-{\bold r})-\delta^{3}({\bold x}) \right\}$ satisfying 
the conservation law $\partial_{\mu}j^{\mu}_{Q}=0$. 
Here $Q$ and $-Q$ are point charges at 
$\bold{x}=\bold{r}$ and $\bold{x}=\bold{0}$, respectively. 
(For a particular configuration of the classical fields 
$\phi_{0}{}^{i}$, the current $j^{\mu}$ could reduce to 
$j^{\mu}_{Q}$. Conversely, in some specific cases, $j^{\mu}$ might be 
expressed in a form of superposition of $j^{\mu}_{Q}$.) 
Substituting $j^{\mu}_{Q}$ into Eqs. (\ref{50}) and (\ref{53})  
and choosing a constant vector $(0,\, \bold{n})$ with the condition 
$\bold{n}/\!/\bold{r}$ to be $n^{\mu}$, we can calculate 
the effective potentials $V_{j}$ and $V_{J}$ defined by 
$-V_{j}\int dx^{4} =W_{j}^{(0)}[j^{\mu}_{Q}]$ and 
$-V_{J}\int dx^{4} =W_{J}^{(0)}[j^{\mu}_{Q}]$.  The result reads 
\begin{eqnarray}
V_{j}(r)&=&-\frac{(\rho_{3}Q/g_{0})^{2}}{4\pi \rho_{1}} 
\frac{e^{-(g(\mu)\varLambda/4\pi)r}}{r} \,, 
\label{54}
\\ 
V_{J}(r)&=& 
\frac{Q^{2}M_{2}^{2}}{8\pi} 
\bigg[ \ln \bigg(1+\frac{\widetilde{\varLambda}^{\,2}}{M_{1}^{2}} 
\bigg) \bigg] r 
\label{55}
\end{eqnarray}
%
up to irrelevant infinite constants. 
Here $r\equiv|\bold{r}|$, and $\widetilde{\varLambda}$ is another ultraviolet 
cutoff. 
(In deriving Eq.~(54) from Eq.~(50), it has been taken into account 
that $j^{0}=-ij^{4}$.) 
Therefore the EAGT allows a composite of the 
Yukawa and the linear potentials, $V_{j}+V_{J}$, even at the tree level. 
This result is quite desirable for describing color confinement 
\cite{Suzuki,SST}. 
Comparing Eq. (\ref{55}) with the linear potential obtained in 
the MAATGT with ATC,\footnote{\setlength{\baselineskip}{6mm} 
The linear potential obtained in the MAATGT with ATC is    
$V_{J}(r)= 
\frac{Q^{2}m^{2}}{8\pi g_{\rm H}^{2}} 
\Big[ \ln \Big(1+\frac{\widetilde{\varLambda}^{\,2}}{m^{2}} 
\Big) \Big] r $. Here $m$ is a constant with dimension of mass and 
$g_{\rm H}$ is a constant identified with the gauge 
coupling constant of the EDAHM. 
In ref. \cite{Deguchi-Kokubo}, $g_{\rm H}=1$ was put for convenience. }
we see that $M_{1}/M_{2}$ corresponds to the gauge coupling constant of 
the extended dual Abelian Higgs model (EDAHM). 
The constants $k_{1}$ and $k_{2}$ could be determined 
via Eqs. (\ref{38}) and (\ref{39}) in such a way that $M_{1}/M_{2}$ 
reproduces the (running) coupling constant of the EDAHM.

Now we should make a comment. In terms of the unite isovector 
$\vec{n}=\vec{n}(\phi_{0})=({n}^{B})$ defined by  
${n}^{B}(\phi_{0})T^{B}=v(\phi_{0})T^{3}v^{-1}(\phi_{0})$, 
the antisymmetric tensor 
$J_{\mu\nu}=\epsilon^{3bc}\tilde{e}_{\mu}{}^{b} \tilde{e}_{\nu}{}^{c}$ 
is written as 
$J_{\mu\nu}=\vec{n}\cdot (\partial_{\mu}\vec{n}\times
\partial_{\nu}\vec{n})
\equiv \epsilon^{BCD}n^{B}\partial_{\mu}n^{C}\partial_{\nu}n^{D}$. 
With this expression, we see that ${\cal F}_{\mu\nu}$ in 
Eq. (\ref{35}) is just the so-called 't Hooft tensor occurring 
in the theory of 't Hooft-Polyakov magnetic monopole \cite{TPM}, provided 
that $n^{B}$ are identified with the Higgs fields normalized 
in isospace. As is known in this theory, the magnetic current 
$k^{\mu}=\partial_{\nu}*{\cal F}^{\nu\mu}
=\partial_{\nu}*{J}^{\nu\mu}$ remains non-vanishing owing to  
non-triviality of the second homotopy group of $SU(2)/U(1)$, i.e., 
$\pi_{2}(SU(2)/U(1))=\pi_{1}(U(1))={\Bbb Z} \cite{Wiese,Coleman}$. 
(Here $*$ indicates the Hodge star operation.) 
Comparing $j^{\mu}=\partial_{\nu}J^{\nu\mu}$ with 
$k^{\mu}=\partial_{\nu}*{J}^{\nu\mu}$, we see that $j^{\mu}$ is certainly 
a color-electric current, not a color-magnetic current. 
For this reason, $W_{j}^{(0)}+W_{J}^{(0)}[j^{\mu}]$ is interpreted as 
a generating functional that describes interaction between 
color-electric currents; accordingly $V_{j}+V_{J}$ is understood to be 
a static potential between color-electric charges.

\section{Remarks}

One may ask here what is the origin of the color-electric current 
$j^{\mu}$. We can give no satisfactory answer to this question now, 
and so we will only mention a tentative idea towards making a correct reply. 
Recall the fact that $\phi_{0}{}^{i}$ is a solution of the Euler-Lagrange 
equation in Eq. (\ref{6}). Since this equation involves 
the gluon fields $A_{\mu}{}^{B}$ in the form of external fields, 
$\phi_{0}{}^{i}$ can be regarded as functionals of $A_{\mu}{}^{B}$,  
i.e., $\phi_{0}{}^{i}=\phi_{0}{}^{i}[A_{\mu}{}^{B}]$, or more physically, 
as (composite) fields that inherit degrees of freedom of the gluon 
fields.\footnote{
\setlength{\baselineskip}{6mm}
Scalar fields similar to $\phi_{0}{}^{i}$, or more precisely 
to $n^{B}$, have been introduced by Ichie and Suganuma in a somewhat 
different context \cite{Ichie-Suganuma}. 
They called those scalar fields 
{\em gluonic Higgs fields}, insisting that the scalar fields are 
analogous to non-Abelian Higgs fields but are composite fields of gluons. 
Unlike Ichie-Suganuma's procedure of introducing the scalar fields, 
we have obtained $\phi_{0}{}^{i}$ as a classical configuration of 
the ^^ ^^ dynamical" NG scalar fields $\phi^{i}$ that are 
identified with the longitudinal modes of the massive off-diagonal gluons.} 
Regarded $\phi_{0}{}^{i}$ as gluonic scalar fields in this sense, 
$J^{\mu\nu}$ and $j^{\mu}$ are interpreted as currents consisting of  
gluonic degrees of freedom. Thus it seems natural to assign the origin of 
the color-electric current $j^{\mu}$ to the gluon fields. If this idea is 
acceptable, $V_{j}+V_{J}$ could be considered as 
a potential that makes a contribution to gluon confinement.

We now focus attention to the leading terms of ${\cal L}_{2}$ in 
Eq. (\ref{49}), including the quartic terms in $\tilde{e}_{\mu}{}^{b}$ 
and the quadratic terms in 
$\widetilde{\nabla}^{\mu bc} \tilde{e}_{\mu}{}^{c}$. 
The sum of the leading terms can be seen from Eqs. (\ref{25}), (\ref{29}), 
(\ref{30}) and (\ref{34}) to be 
\begin{eqnarray}
{\cal L}_{2}^{(0)}&=& \frac{\varLambda^{2}}{32\pi^{2}} 
\tilde{e}_{\mu}{}^{b}\tilde{e}^{\mu b}
-{1\over{4g_{0}^{2}}}\bigg(\rho_{4}+\frac{g_{0}^{2}}{32\pi^{2}}
\ln \frac{\mu}{\varLambda} \bigg) J_{\mu\nu}J^{\mu\nu} 
\nonumber \\ 
& &
+{1\over128\pi^{2}} \ln \frac{\mu}{\varLambda}
(\tilde{e}_{\mu}{}^{b}\tilde{e}^{\mu b})^{2} 
-\frac{\rho_{2}}{2\alpha_{0}g_{0}^{2}} 
\widetilde{\nabla}^{\mu bc} \tilde{e}_{\mu}{}^{c} 
\widetilde{\nabla}^{\nu bd} \tilde{e}_{\nu}{}^{d} \,,
\label{56}
\end{eqnarray}
%
where 
\begin{eqnarray}
\rho_{4}\equiv 1-{4\over k_{2}}-g_{0}^{2}
\bigg({17\over{96\pi^{2}}}+{k_{1}\over{2\pi^{2}k_{2}^{2}}}\bigg)
\ln \frac{\mu}{\varLambda} \,.
\label{57}
\end{eqnarray}
%
In terms of $n^{B}$, ${\cal L}_{2}^{(0)}$ is written as 
\begin{eqnarray}
{\cal L}_{2}^{(0)}&=&
\frac{\varLambda^{2}}{32\pi^{2}} \partial_{\mu}\vec{n}\cdot
\partial^{\mu}\vec{n} 
-\frac{\rho_{4}}{4g_{0}^{2}}
(\partial_{\mu}\vec{n}\times\partial_{\nu}\vec{n})\cdot
(\partial^{\mu}\vec{n}\times\partial^{\nu}\vec{n})
\nonumber \\ 
& &+{1\over128\pi^{2}}\ln\frac{\mu}{\varLambda}
(\partial_{\mu}\vec{n}\cdot\partial_{\nu}\vec{n})
(\partial^{\mu}\vec{n}\cdot\partial^{\nu}\vec{n}) 
\nonumber \\ 
& &
-\frac{\rho_{2}}{2\alpha_{0}g_{0}^{2}}
(\vec{n}\times \square\vec{n})\cdot(\vec{n}\times \square\vec{n}) \,.
\label{58}
\end{eqnarray}
%
This is certainly the Lagrangian that defines a generalized Skyrme-Faddeev 
model (GSFM) \cite{Niemi-Cho}. 
Therefore it is concluded that the EAGT involves a GSFM. 
As is seen from Eq. (\ref{45}), $\rho_{2}$ vanishes with the choice of 
gauge parameter 
$\alpha_{0}=-16\pi^{2}g_{0}^{-2}[\,\ln(\mu/\varLambda)]^{-1}$. 
In this case, ${\cal L}_{2}^{(0)}$ reduces to the Lagrangian  
presented by de Vega to investigate closed-vortex configurations \cite{Vega}. 
If the last two terms in Eq. (\ref{58}) are removed, 
${\cal L}_{2}^{(0)}$ agrees with the Lagrangian of the Skyrme-Faddeev model 
\cite{Skyrme-Faddeev}.  Faddeev and Niemi have conjectured that 
the Skyrme-Faddeev model would be appropriate for 
describing $SU(2)$ YM theory in the low energy limit \cite{Faddeev-Niemi}. 
If their conjecture is true, $n^{B}$ should be considered as 
the fields that realize gluonic degrees of freedom relevant to the low energy 
limit. This interpretation of $n^{B}$ appears consistent 
with our statement on $\phi_{0}{}^{i}$ made above.

Until now the ultraviolet cutoff $\varLambda$ remains a free parameter; 
we briefly discuss how to fix $\varLambda$. 
First we recall that ${\cal L}_{\phi}$ in Eq. (\ref{2}) was expanded 
about the classical fields $\phi_{0}{}^{i}$ in powers of $\sigma^{i}\,$: 
\begin{eqnarray}
{\cal L}_{\phi}=\frac{m_{0}^{2}}{2g_{0}^{2}} 
\tilde{e}_{\mu}{}^{b} \tilde{e}^{\mu b} 
+\cdots \,. 
\label{59}
\end{eqnarray}
%
Here the relation  
$g_{ij}(\phi_{0})\partial_{\mu}\phi_{0}{}^{i} \partial^{\mu}\phi_{0}{}^{j}
=\tilde{e}_{\mu}{}^{b} \tilde{e}^{\mu b}$ has been used. 
From a viewpoint of the BRST formalism, it can be understood that 
because of the BRST symmetry, 
the quadratic divergences due to quantum effects of 
$\phi^{i}$, $c^{a}$ and $\bar{c}^{a}$ cancel out in calculating 
quantum correction for the classical term 
$(m_{0}^{2}/2g_{0}^{2}) \tilde{e}_{\mu}{}^{b} \tilde{e}^{\mu b}$. 
(This fact can be seen by comparing Eq. (\ref{9}) with Eq. (\ref{34}).) 
For this reason, the classical term is not affected by quantum effects 
and will remain in the EAGT without any change. 
We now notice that the Lagrangian ${\cal L}_{\rm tot}$ includes 
$(\varLambda^{2}/32\pi^{2})\tilde{e}_{\mu}{}^{b} \tilde{e}^{\mu b}$, 
see Eq. (\ref{34}). 
In the EAGT, this is the only term that takes the same form as 
the classical term. 
Thus it is natural to identify 
$(\varLambda^{2}/32\pi^{2})\tilde{e}_{\mu}{}^{b} \tilde{e}^{\mu b}$ 
with the classical term by imposing the condition 
$\varLambda^{2}/32\pi^{2}=m_{0}^{2}/2g_{0}^{2}\,$, which fixes 
$\varLambda$ at 
\begin{eqnarray}
\varLambda_{0}\equiv \frac{4\pi m_{0}}{g_{0}} \,.
\label{60}
\end{eqnarray}
%
Even if $m_{0}$ is not so large value, 
the cutoff $\varLambda_{0}$ will be regarded as large 
so long as $g_{0}/4\pi$ is sufficiently small.  
In such a case, the perturbative treatment of the 
higher order and/or higher derivative terms in ${\cal L}_{\rm tot}$ 
makes good sense.

\section{Conclusions}

We have derived an EAGT of the SKG formalism 
by following a method of Abelian projection. 
There the off-diagonal gluon fields involving longitudinal modes were 
treated as fields that produce quantum effects on the diagonal gluon 
fields and other fields relevant at a long-distance scale.  
In deriving the EAGT, we have employed, 
instead of the MA gauge fixing term used in earlier papers, 
an appropriate gauge fixing term in Eq. (\ref{18}) 
which breaks the {\em local} $SU(2)$ gauge symmetry, 
while preserves the {\em global} $SU(2)$ gauge symmetry. 
(An equation corresponding to the MA gauge condition was obtained 
as the Euler-Lagrange equation for the NG scalar fields, 
see Eq. (\ref{6}).) 
This choice of gauge fixing term guarantees the conservation of color 
charge and has made evaluating the ghost sector simple.

We have shown that the EAGT involves a gauge fixed version 
of the MAATGT with ATC, allowing a composite of the Yukawa and 
the linear potentials at the tree level of the EAGT. 
We can not immediately identify this linear potential with the confinement 
potential, because it was derived via a perturbative procedure 
together with the one-loop approximation. 
However, we could say that the ^^ ^^ germ" of confinement potential was found. 
The composite potential derived here is understood to be 
a static potential between color-electric charges.  
It was also pointed out that the origin of these charges may be assigned 
to the gluon fields. For this reason, the composite potential could make 
a contribution to confinement of gluons.  
Finally we have seen that the EAGT involves the Skyrme-Faddeev 
model in addition to the MAATGT with ATC.

In the present letter we have dealt only with the three terms 
$W_{j}^{(0)}$, $W_{J}^{(0)}$ and ${\cal L}_{2}^{(0)}$ among the terms 
included in Eq. (\ref{49}). 
Some of the remaining terms will produce correction for the potential 
$V_{j}+V_{J}$. Since this potential corresponds to the static 
potential found in the EDAHM in the London limit, such correction 
should be identified with variation from the London limit 
\cite{Kondo 2,Kondo 3}. 
We then expect that the scalar field $\varPhi$ plays a role of the 
residual Higgs field in the EDAHM, with choosing suitable values for  
the constants $k_{1}$ and $k_{2}$. 
Anyway, investigation of the correction is important for comparing the EAGT 
with the EDAHM. 
Besides clarifying the mass-generation mechanism of off-diagonal gluons, 
we hope to address this issue in the future.

\begin{acknowledgements}
We are grateful to members of the Theoretical 
Physics Group at Nihon University for their encouragements. 
We would like to thank Dr. B. P. Mandal for a careful reading of 
the manuscript.
\end{acknowledgements}

\end{document}